\documentclass[fleqn,10pt]{wlscirep}
\usepackage[utf8]{inputenc}
\usepackage[T1]{fontenc}
\usepackage{siunitx}
\usepackage{amsmath}
\usepackage{xparse,nameref}
\usepackage{hyperref}
\title{Laue three dimensional neutron diffraction}

\author[1,2,*]{Marc Ravent\'os}
\author[3]{Michael Tovar}
\author[1]{Marisa Medarde}
\author[1,4]{Tian Shang}
\author[1,5,6]{Markus Strobl}
\author[5]{Stavros Samothrakitis}
\author[1]{Ekaterina Pomjakushina}
\author[1]{Christian Gr\"{u}nzweig}
\author[7,*]{S{\o}ren Schmidt}

\affil[1]{Paul Scherrer Institut, Switzerland}
\affil[2]{University of Geneva, Switzerland}
\affil[3]{Helmholtz Zentrum Berlin, Germany}
\affil[4]{\'Ecole Polytechnique F\'ed\'erale de Lausanne, Switzerland}
\affil[5]{Nuclear Physics Institute of the CAS, Czech Republic}
\affil[6]{Niels Bohr Institute, Denmark}
\affil[7]{Technical University of Denmark, Denmark}
\affil[*]{marc.raventos@psi.ch, ssch@fysik.dtu.dk}

\begin{abstract}
This article presents a measurement technique and data analysis tool to perform 3D grain distribution mapping and indexing of oligocrystalline samples using neutrons: Laue three-dimensional neutron diffraction (Laue3DND). The approach builds on forward modelling used for correlation and multiple fitting of the measured diffraction spots relative to individual grains. This enables not only to identify individual grains, but also their position and orientation in the sample. The feasibility and performance of the Laue3DND approach are tested using multi-grain synthetic datasets from cubic ($\alpha$-Fe) and tetragonal (YBaCuFeO$_5$) symmetries. Next, experimental results from two data sets measured at the FALCON instrument of Helmholtz-Zentrum Berlin are presented: A cylindrical alpha iron ($\alpha$-Fe) reference sample with 5 mm diameter and 5 mm height, as well as a 2 mm$^3$ layered perovskite (YBaCuFeO$_5$). Using Laue3DND, we were able to retrieve the position and orientation of 97 out of 100 grains from a synthetic $\alpha$-Fe data set, as well as 24 and 9 grains respectively from the $\alpha$-Fe and YBaCuFeO$_5$ sample measured at FALCON. Results from the synthetic tests also indicate that Laue3DND is capable of indexing 10 out of 10 grains for both symmetries in two extreme scenarios: using only 6 Laue projections and using 360 projections with extremely noisy data. The precision achieved in terms of spatial and orientation resolution for the current version of the method is 430 $\mu$m and \ang{1} respectively. Based on these results obtained, we are confident to present a tool that expands the capabilities of standard Laue diffraction, providing the number, position, orientation and relative size of grains in oligocrystalline samples.

\end{abstract}
\begin{document}

\flushbottom
\maketitle
%
%
\thispagestyle{empty}


\section*{Introduction}

Understanding the link between a material structure at different length scales and its emerging macroscopic properties is a general theme of material science. In the case of crystalline materials, retrieving 3D spatial information about the individual crystallites (grains) in the bulk non-destructively has been the motivation behind several methods developed with X-rays and widely applied for corresponding valuable studies since the millennium \cite{Poulsen2001,Larson2002,Jensen2006,Ludwig2008,Johnson2008}. These methods make possible the retrieval of grain maps from millimetric samples with sub-micron resolution. Sub-micron resolutions are outside the reach of neutron grain mapping techniques, but neutrons bear the promise to retrieve grain maps with sub-millimeter resolution of larger sample volumes due to their better penetration characteristics for many structural engineering materials and dense crystals. Grain mapping of bulky engineering samples is particularly interesting for large directionally grown pieces, such as nickel-based turbine blades, and large samples undergoing phase transformations during use, such as iron-based shape memory alloys. In the field of solid-state physics, a 3D grain mapping characterization tool with neutrons enables the utilization of imperfect crystals for diffraction studies, (e.g.) by providing 3D information about the position of the largest grain within a sample so it can be cut out. In both metallurgy and solid-state physics, neutrons are better suited than X-rays for \textit{in-situ} testing with bulkier sample environments. A first approach and proof-of-principle experiment with neutrons (nDCT) was reported recently \cite{Peetermans2014}, using full sample illumination and a neutron beam with a narrow energy spectrum. The result was a grain boundary topological 3D reconstruction of 13 grains from an aluminum sample, measured with cold neutrons at the ICON beamline \cite{Kaestner2011} of Paul Scherrer Institut (PSI). More recently, a substantially more efficient Time-of-Flight (ToF) approach utilizing a pulsed neutron source has been introduced\cite{Cereser2017a} utilizing the SENJU beamline \cite{Tamura2012} at J-PARC (ToF3DND) in conjunction with a timepix transmission imaging detector \cite{Tremsin2011}. ToF3DND enabled indexing and reconstruction of more than a hundred grains under full sample illumination. Both realizations of grain mapping with neutrons were based on partial or full wavelength resolved measurements enabling diffraction analyses and retrieval of 3D grain maps. Therefore, introducing a white beam Laue diffraction technique facilitates grain mapping at most neutron sources, in particular at continuous sources, where energy selection implies selecting only a part of available flux. Thus we present Laue3DND, which draws partially on intense continuous white beam flux but on the other hand on complex computational efforts in a forward modelling approach.

Firstly, the necessary crystallographic and geometrical concepts to build the \textbf{forward model} are explained, structured in the \textbf{sample, laboratory and detector reference systems}. Once the Laue patterns can be simulated, the \textbf{solver} is introduced and structured in \textbf{seeding, single grain fitting and global fitting}. These are the set of iterative algorithms implemented to find the best possible match between the simulated and measured spots. The indexing procedure is then finished, so the \textbf{analysis} of the diffracted intensities can be carried out. In order to test the \textbf{performance} of Laue3DND a series of tests are conducted using synthetic data sets, which provide benchmarks for robustness, precision and limitations of the method. Next, the \textbf{experimental setup} of the FALCON beamline of Helmholtz Zentrum Berlin (HZB) is detailed, followed by the \textbf{experimental results} from the \textbf{$\alpha$-Fe and YBaCuFeO$_5$ oligocrystalline samples}. Finally, the results obtained, the current performance of the method and the future challenges and improvements are laid out in the \textbf{discussion and conclusions}.  


\section*{Forward model}

\begin{figure}[hbt]
\centering
\includegraphics[scale=0.8]{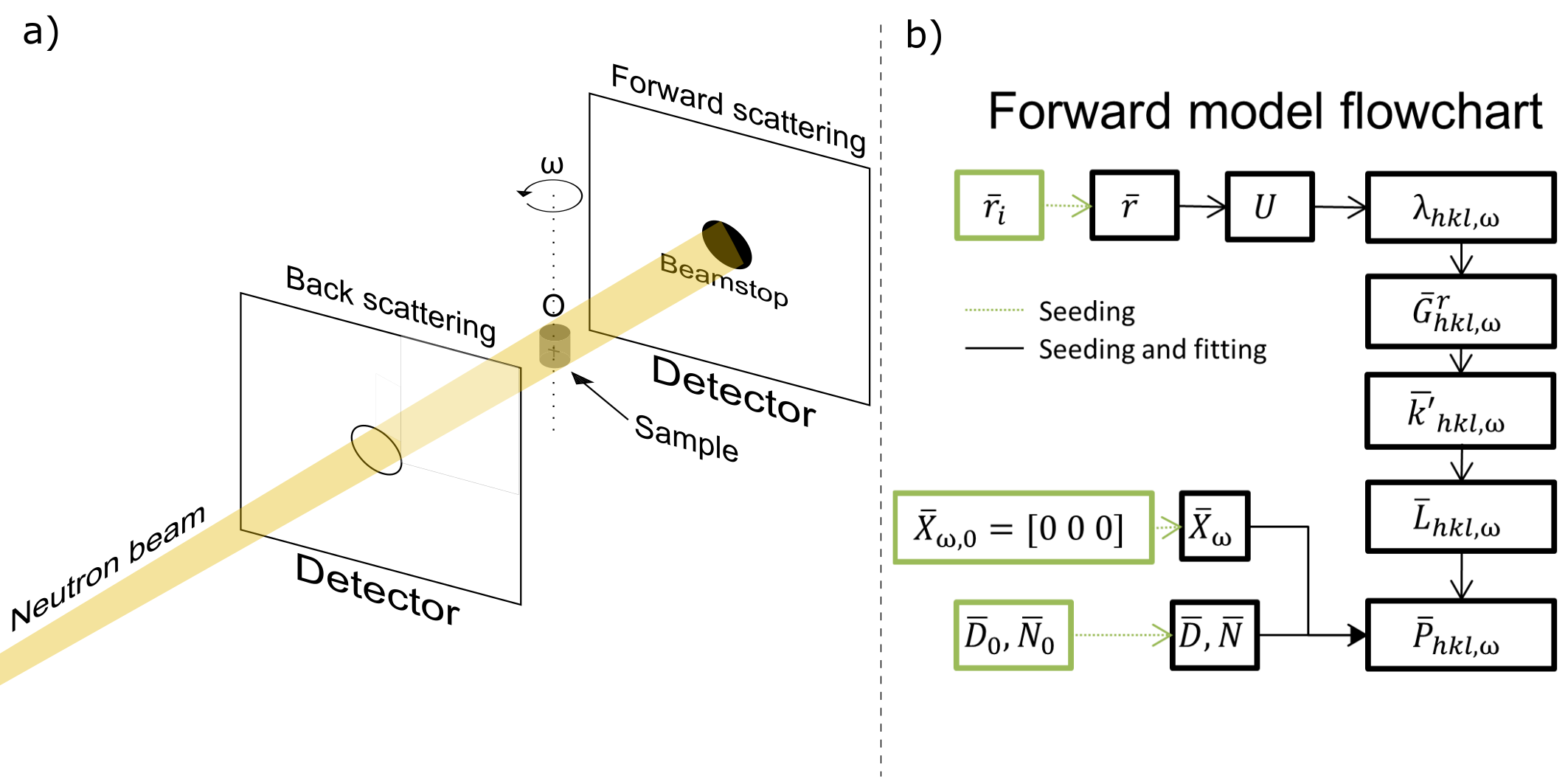}
\caption{a) Scheme of a double detector Laue setup. The neutron beam traverses the center of the back scattering detector through hole in the back scattering detector and meets the beam stop in front of the forward scattering detector. Ideally, the sample is placed in the center of the rotation axis $\omega$, as shown in the figure. b) Flowchart of the forward model. From a given grain orientation $\bar{r}$, the valid scattering wavelengths $\lambda$ are calculated, followed by the reciprocal scattering vector $\bar{G^r}$ and the direction of the diffracted vector $\bar{k'}/\bar{|k'|}$ which is $\bar{L}$ in the sample reference system. The position of the grain $\bar{X}$ and the detector position $\bar{D}$ and its orientation $\bar{N}$ are then used together with $\bar{L}$ to calculate the position of diffraction spot $\bar{P}$. This process is repeated for every $hkl$ plane and $\omega$ rotation step. The black labeled part of the flowchart (Seeding and fitting) is run every time the forward model is used, while the green part is only used during the first part of the algorithm (Seeding).}
\label{Fig:FALCON}
\end{figure}

The forward model is the tool that allows simulating diffraction patterns from crystal and beamline parameters as they would be during an experiment. All the crystallographic and geometrical calculations are performed in the forward model as shown in Fig. \ref{Fig:FALCON}, so later the solver can compare the position of the predicted spots ($\bar{P}$) with the measured spots.

Ultimately, the forward model simulates the Laue diffraction pattern measured at one or more given detectors ($\bar{D}$ and $\bar{N}$), for a given grain orientation ($\bar{r}$) and a given position of the grain in space ($\bar{X}$). In order to do this, three different reference systems are used throughout the forward modelling: the sample, laboratory and detector reference systems.

\subsection*{Sample reference system}

\begin{figure}[hbt]
\centering
\includegraphics[scale=0.9]{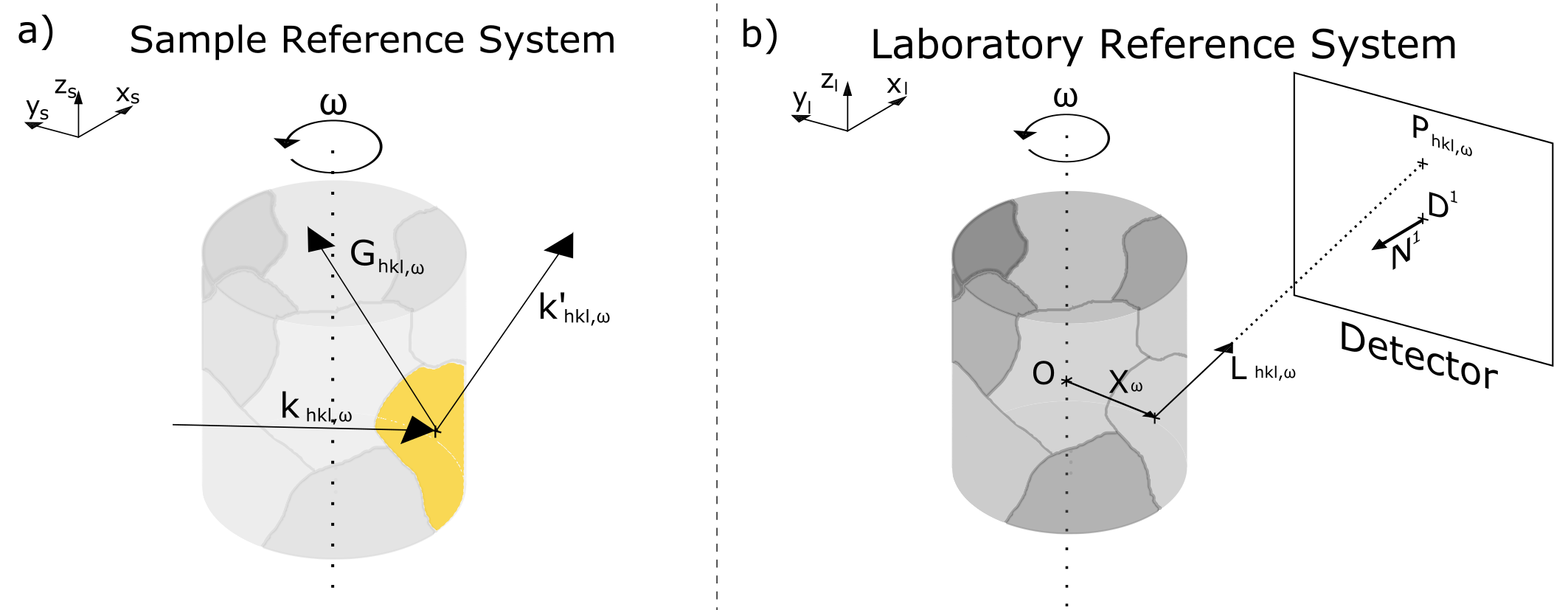}
\caption{a) Scheme of the sample reference system, where $\bar{k}_{hkl,\omega}$ represents the direction of the incoming beam and $\bar{k'}_{hkl,\omega}=\bar{G}_{hkl,\omega}+\bar{k}_{hkl,\omega}$ is the direction of the diffracted beam and $\bar{G}_{hkl,\omega}$ is a reciprocal lattice vector for the plane $hkl$ at a given rotation angle $\omega$. b) Scheme of the laboratory reference system.  Given the detector position and orientation ($\bar{D}, \bar{N}$), the diffraction spot $\bar{P}_{hkl,\omega}$ can be calculated for every $\omega$, as the intersection of the line formed by the scattering vector $\bar{L}_{hkl,\omega}$ and the grain position $\bar{X}_{\omega}$ with the detector plane.}


\label{Fig:SRS_LRS}
\end{figure}

The sample reference system (SRS) is used to calculate the direction of the diffracted vector ($\bar{k'}_{hkl,\omega}/|\bar{k'}_{hkl,\omega}|$) with respect to the incoming beam ($\bar{k}_{hkl,\omega}/|\bar{k}_{hkl,\omega}|$), for every $hkl$ plane and wavelength ($\lambda_{hkl,\omega}$) combination that satisfies the Bragg condition for every rotation angle ($\omega$). To simplify the notation we use  

\begin{equation}
\centering
\bar{L}_{hkl,\omega}^{samp}=\bar{k'}_{hkl,\omega}/|\bar{k'}_{hkl,\omega}|,
\label{eq:L_k}
\end{equation}

where $\bar{L}_{hkl,\omega}^{samp}$ is the unit vector with the direction of the diffracted beam in the SRS. A scheme of the sample reference system is shown in Fig. \ref{Fig:SRS_LRS}a. 

First, the forward model requires a grain orientation defined in the sample reference system. We define the orientation of the grain as $\bar{r}$, a three component vector in Rodrigues space\cite{He2007}. Defining the orientation in Rodrigues formulation has two main benefits with respect to Euler angles: any orientation can be defined with only three components (which simplifies the optimization process) and the orientations are uniquely defined, given that the $\bar{r}$ vector lies in the fundamental zone of the given crystal symmetry.
The Rodrigues vector $\bar{r}$ is used to calculate the rotation matrix of the grain orientation by


\begin{equation}
\centering
U=\frac{1}{1+r_1^2+r_2^2+r_3^2}
\begin{bmatrix}
1+r_1^2-r_2^2-r_3^2 & 2\cdot(r_1\cdot r_2 + r_3) & 2\cdot(r_1\cdot r_3 - r_2) \\
2\cdot(r_1\cdot r_2 - r_3) & 1-r_1^2+r_2^2-r_3^2 & 2\cdot(r_2\cdot r_3 + r_1) \\
2\cdot(r_1\cdot r_3 + r_2) & 2\cdot(r_2\cdot r_3 - r_1) & 1-r_1^2-r_2^2+r_3^2\\
\end{bmatrix},
\label{eq:U}
\end{equation}

where $U$ is the rotation matrix of the grain orientation and $r_1$, $r_2$ and $r_3$ are the three components of $\bar{r}$. 
Once the orientation of the grain is defined, the next step is to calculate all the neutron wavelengths which can potentially fulfill the Bragg condition by

\begin{equation}
\centering
 \lambda_{hkl,\omega}= -\frac{4\pi}{|B\bar{G}_{hkl}|^2} \left(\Gamma_\omega UB \bar{G}_{hkl}\right)_1,
\label{eq:lambda}
\end{equation}

where $B$ is the transformation matrix between the Cartesian lattice and the reciprocal space, $\bar{G}_{hkl}$ are the Miller indices of the reflecting $hkl$ planes, $\Gamma_\omega$ is the right-hand rotation matrix around the z-axis for the angle $\omega$, $U$ is the rotation matrix for the grain orientation, and $\lambda_{hkl,\omega}$ is the resulting diffracted wavelength. With the formulation given in equation \eqref{eq:lambda} only the first component of the product $\Gamma_\omega UB \bar{G}_{hkl}$ is used for the calculation, as explained in Ref. \citen{Schmidt2014}. The next step in the forward model is to obtain all the reciprocal lattice vectors for the given structure, orientation and diffraction planes, which satisfy the Bragg condition.
From the neutron wavelengths found in equation \eqref{eq:lambda}, we remove those which are not present in the incident neutron spectrum in order to find the valid reciprocal scattering vectors, by

\begin{equation}
\centering
\bar{G}^{r}_{hkl,\omega}=\frac{\lambda_{hkl,\omega}}{4\pi}UB\bar{G}_{hkl},
\label{eq:Gr}
\end{equation}

 where $\bar{G}^{r}_{hkl,\omega}$ are the reciprocal scattering vectors. Finally, one can calculate the direction of the diffracted beam for each reciprocal scattering vector by

\begin{equation}
\centering
\bar{L}_{hkl,\omega}^{samp}=2\bar{G}^{r}_{hkl,\omega}+\Gamma_\omega^{-1}\mid_1,
\label{eq:L_samp}
\end{equation}

where $\Gamma_\omega^{-1}\mid_1$ is the first column of the inverse of $\Gamma_\omega$. For details on the formulation presented on equations \eqref{eq:lambda}, \eqref{eq:Gr} and \eqref{eq:L_samp} refer to Refs. \citen{Cereser2017a,Schmidt2014}.

\subsection*{Laboratory reference system}

The laboratory reference system (LRS) is used to calculate the intersection point ($\bar{P}_{hkl,\omega}$) with the detector of the line formed by the diffraction vector ($\bar{L}_{hkl,\omega}$) and the center-of-mass (CMS) of the grain ($X_\omega$). That intersection point is the position of the diffraction spot on the detector in the LRS.
We first transform $\bar{L}_{hkl,\omega}^{samp}$ into de LRS by

\begin{equation}
\centering
\bar{L}_{hkl,\omega}^{lab}=\Gamma_\omega\bar{L}_{hkl,\omega}^{samp},
\label{eq:L_lab}
\end{equation}

where $\bar{L}_{hkl,\omega}^{lab}$ is the direction of the diffracted beam in the LRS, as shown in Fig. \ref{Fig:SRS_LRS}b.


The position of the CMS of the grain ($X_\omega$) is initially assumed to be at the origin of the coordinate system ($O$) as a first approximation. The position of the predicted diffraction spot $\bar{P}^{lab}_{hkl,\omega}$ on the detector, given by


\begin{equation}
\centering
\bar{P}^{lab}_{hkl,\omega}=\bar{X}_{\omega} + t\bar{L}_{hkl,\omega}^{lab},
\label{eq:P}
\end{equation}

where $\bar{X}_{\omega}$ is the CMS of the grain in the LRS  and 

\begin{equation}
\centering
t=\frac{\bar{N}\cdot(\bar{D}-\bar{X}_{\omega})}{\bar{N}\cdot\bar{L}_{hkl,\omega}^{lab}},
\label{eq:t}
\end{equation}

where $t$ is the modulus required for the unit vector $\bar{L}_{hkl,\omega}^{lab}$ to reach the detector from $X_\omega$,  $\bar{D}$ is the detector position and $\bar{N}$ is the orientation. 

The orientation of the detector plane is given by

\begin{equation}
\centering
\bar{N}_f=A_f
\begin{bmatrix}
-1 \\
0 \\
0 \\
\end{bmatrix}
\centering
\label{eq:N_f}
\end{equation}

and

\begin{equation}
\bar{N}_b=A_b
\begin{bmatrix}
1 \\
0 \\
0 \\
\end{bmatrix},
\label{eq:N_b}
\end{equation}

where $A_f$ and $A_b$ are 3D right hand rotation matrices for the forward \eqref{eq:N_f} and backward \eqref{eq:N_b} detectors respectively, which account for the three tilt directions of the detectors.
Following this procedure for each of the detectors the position $\bar{P}^{lab}_{hkl,\omega}$ of every diffraction spot is obtained for every $hkl$, $\omega$ and a given orientation.

\subsection*{Detector reference system}

To calculate the difference between experiment and calculation the modelled diffraction spots need to be  projected from the grain in the LRS onto the detector in the detector reference system (DRS) by

\begin{equation}
\centering
\bar{P}^{det}_{hkl,\omega}=A^{-1}(\bar{P}_{hkl,\omega}^{lab}-\bar{D}),
\label{eq:Pdet}
\end{equation}

where $\bar{P}^{det}_{hkl,\omega}$ is the diffraction spot position in the DRS, after subtracting the position of the detector $\bar{D}$ from $\bar{P}^{lab}_{hkl,\omega}$ and applying the inverse rotation matrix $A^{-1}$ from \eqref{eq:N_f} for the forward detector or \eqref{eq:N_b} for the backward detector.

\begin{figure}[hbt]
\centering
\includegraphics[scale=0.9]{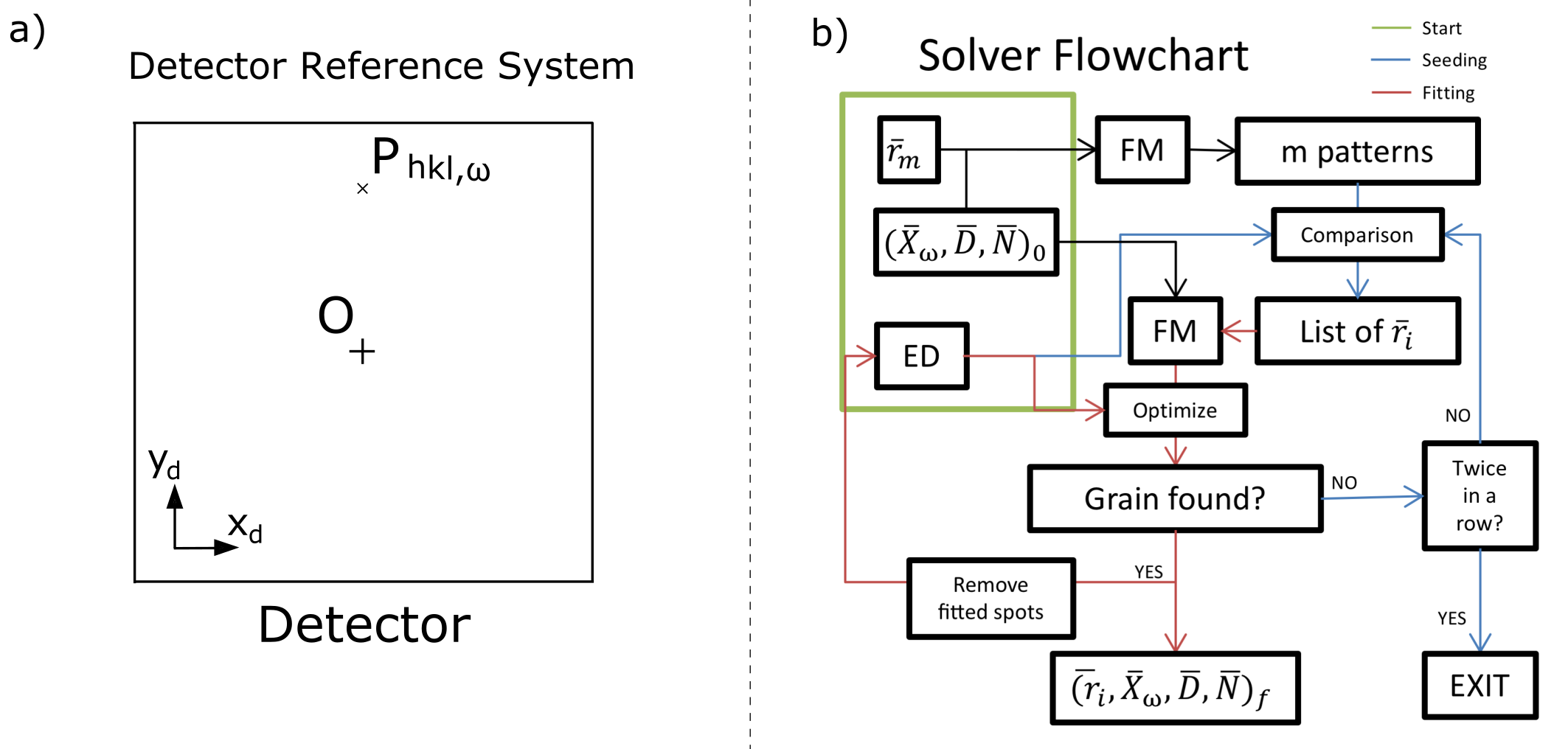}
\caption{a) Detector reference system. $O$ is the center of the detector and the origin of coordinates, and $\bar{P}_{hkl,\omega}$ is the position of the diffraction spot. b) Flowchart of the solver algorithm. FM stands for Forward Model, and ED for Experimental Data. The blue lines describe the seeding process, while the red lines describe the fitting process. Both algorithms are iterated during the solver execution to find the optimal parameters for every grain fit. The green label indicates the starting values for the solver.}
\label{Fig:DRS_SolverFC}
\end{figure}

Once the diffraction spot position is calculated in the DRS ($\bar{P}^{det}_{hkl,\omega}$ in Fig. \ref{Fig:DRS_SolverFC}a), a comparison with the positions of the experimental diffraction spots (peaks) can be performed.

\section*{Solver}\label{sec:TheSolver}

The solver (Fig. \ref{Fig:DRS_SolverFC}b) comprises of the set of algorithms dedicated to find the best possible fit between the experimental data and the grain-by-grain modelled solutions.
The core of the solver is an algorithm which attempts to assign each calculated spot to the CMS of every segmented peak. We argue that a k-nearest-neighbors (KNN) algorithm is the most efficient assignment solution for this task, despite providing non-unique assignments for many spots\cite{Cover1967}. Classic assignment algorithms, like the Hungarian or Munkres\cite{Munkres1957} algorithm, provide a unique assignment but the complexity of these is $O(n^{3})$, while the KNN has a complexity of $O(nd)$ where n is the number of spots and d is the dimensionality of the position (in this case d=2). In the case of an experimental spot being assigned to several predicted spots (or vice versa) we discard all assignments except the one with minimum Euclidean distance. 
Once the KNN is finished, a cost function is computed by using the Euclidean distance of every assignment. Since the experimental data include diffraction spots from many different grains, the algorithm has to be robust enough to identify and optimize a correct orientation with a large number of outliers. Different strategies are followed in order to deal with incorrectly fitted spots, but the underlying principle is to give a higher weight to assignments with the lower Euclidean distance over those with larger Euclidean distances.
\subsection*{Seeding}

Seeding is the overall search of the orientation space in order to find orientations which have higher probability of being close to a grain's orientation in the sample. As explained in Fig. \ref{Fig:DRS_SolverFC}b, the solver first generates a number $m$ of diffraction patterns according to $m$  corresponding divisions of the Rodrigues\cite{He2007} orientation space. The value of $m$  is a compromise between computation time and the size of the fundamental zone. For our experiments, the value of m ranges between 15000 and 45000, which corresponds to a maximum step-width in orientation space between of \ang{3.8} and \ang{2.7} (for a cubic system).
Initially, the forward model assumes a grain position at the origin of the laboratory reference system, and the detectors’ position assumptions are based on approximate distance measurements when configuring the experiment. The resulting diffraction patterns calculated on such detector planes are then compared to the experimental . 
The $m$ calculated orientations are then sorted by the median Euclidean distance between the peaks and the predicted spots. The median of the distances proves to be useful in this case because the goal is to find best suited orientation candidates that have a large number of low-distance assignments without being affected by long-distance outliers, which would be the case if (e.g.) the mean distance was used. Once the $m$ orientations have been sorted by their median distance, the best match is selected for further fitting.

\subsection*{Single grain fitting}\label{ssec:SingleGrainFitting}

A constrained optimization algorithm is used initially to fit each one of the individual grains, in contrast to the seeding process where only direct comparison was performed to sort the orientations by the median. The variables are the detectors' positions and orientations ($\bar{D}$, $\bar{N}$) and the grain's position and orientation ($\bar{X}_\omega$, $\bar{r}$) constituting 6 input parameters for the grain plus 6 parameters per detector used. Since we are using two detectors in our experiments every single grain fit involves a total of 18 variables. The constrains for $\bar{D}$ and $\bar{N}$ are defined by the user and relate to measurement errors. In the cases presented here the tolerances are 10 mm detector misalignment in every direction and \ang{2} for every tilt. With respect to $\bar{X}_\omega$, a 3D space larger than the volume of the sample is set as boundary condition. The best orientation obtained during the seeding process is chosen ($\bar{r}_i$) for the first iteration of the optimization. The volume of the voxel in Rodrigues space which serves as boundary condition for the optimization is defined by the user. In the case of a cubic system we use a volume of 0.066x0.066x0.066 in Rodrigues' space, which corresponds approximately to \ang{7.55} for every Euler angle. Once the boundary conditions for all the parameters are set and the KNN assigns the predicted spots to every peak, we need to estimate the goodness of the fit.
Instead of using the median of the euclidean distances, the cost function used in this case is 

\begin{equation}
C=\frac{N}{
\sum_{j=1}^{N}\frac{1}{(dist_{j}+0.25)}},
\label{eq:Cost}
\end{equation}
where $j$ is the index of $N$ total assignments and $dist_j$ is the euclidean distance of the $j^{th}$ assignment in millimeters. The value of 0.25 added to the distance corresponds to the thickness of the scintillator layer of the detector (0.25 mm), which is a good estimate of the resolution of the detector system. By adding 0.25 to the cost function we give a similar weight to all the assignments which have a distance smaller than the resolution of the detector system.

\begin{figure}[hbt]
\centering
\includegraphics[scale=0.95]{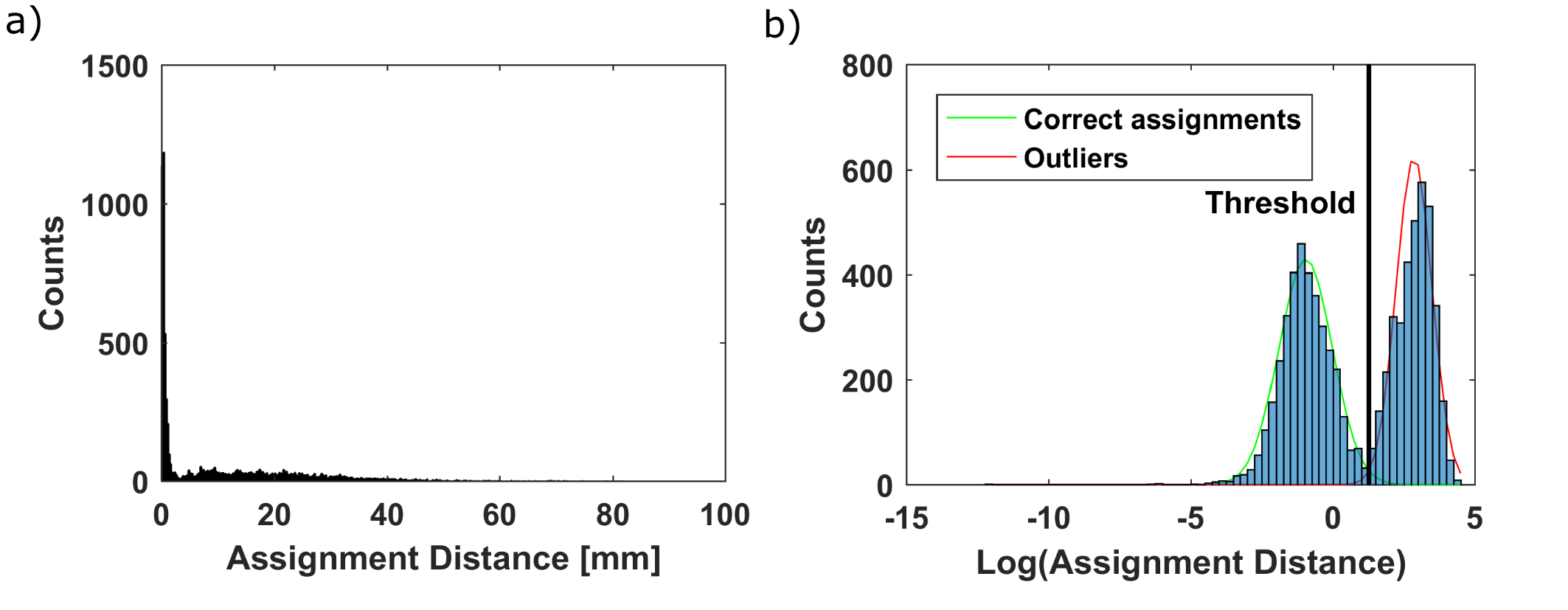}
\caption{a) Histogram of the euclidean distances for the assignment between experimental and predicted spots. b) Histogram of the logarithm of those distances, with Gaussian mixture fit and segmentation of outliers.}
\label{Fig:Histogram}
\end{figure}

Once the cost function is minimized through the constrained optimization algorithm, a criterion to segment the correctly fitted spots from the outliers is applied. Plotting a histogram of the distances resulting from all assignments produces a right-skewed histogram with a long tail, as shown in Fig. \ref{Fig:Histogram}. This is the result of optimizing with the chosen cost function (equation \eqref{eq:Cost}), which gives a higher weight to assignments with small distances and quickly lowers the weight to the assignment when distance increases. This gives the algorithm a higher incentive on reducing the distance of the best assignments even more, while not being affected significantly by an increase in the distance of a \textit{bad} assignment. \textit{Bad} assignments are predicted diffraction spots that do not correlate to corresponding measured peaks. This happens because planes with low structure factors or diffracting neutron wavelengths with low intensity in the incident beam might not provide enough diffracted intensity for the peak to be segmented successfully by the watershed algorithm.

Two overlapping distributions which can be described using a Gaussian fit arise when applying the logarithm to the assignments' euclidean distances, as shown in Fig. \ref{Fig:Histogram}. This approach is not only used for defining the threshold and hence removal of \textit{bad} assignments but also for validation of the orientation fit. Once the logarithms of the assignments' distances is calculated, the algorithm checks if a combination of two Gaussian curves can be fitted to it. The distribution including the lower distances is expected to be the correctly assigned, while the distribution including the higher distances is expected to be the outliers. If such Gaussian mixture model converges, the median of the \textit{good} assignment distances is checked to be lower than a threshold value chosen by the user. We consider that a grain was found if there are more than three diffraction spots assigned per angle, and the median of the assignments' distance is lower than 2 mm. If only one Gaussian distribution can be fitted, the grain is accepted as correct if it fulfills the same conditions regarding number of spots per angle and median of the assignments' distance. Once a grain has been identified as valid, the output parameters, $\bar{D}$, $\bar{N}$, $\bar{X}_\omega$ and $\bar{r}$ are stored, and the peaks correctly assigned are removed from the list of peaks to be fitted.
The fitting process is then repeated for the next grain starting with the next best median distance obtained from the seeding process, until no seed orientation fulfills the described fitting criterion.\\
Then the seeding process starts again creating a new list of $m$ median distances with the peaks yet to be assigned to a grain. If there is no grain found within the first fitting iteration afterwards, the search finishes as seen in Fig. \ref{Fig:DRS_SolverFC}b.

\subsection*{Global fitting}

Global fitting refines all the grain and detector parameters found during the single grain fitting procedure. Diffraction spots from all the predicted grains are generated using the orientations and positions found during the single grain fitting, and are optimized simultaneously by comparison with the peaks through a constrained optimization algorithm. The global fitting algorithm uses again 6 variables per detector plus 6 variables per grain, which in the case of the $\alpha$-Fe sample led to a constrained optimization function with 156 variables. The global fitting creates a competition between the predicted grains to have the best fit for all diffraction spots simultaneously, therefore some of the assigned measured spots are moved to a different grain than the initial one.
The output of this process is the final result of the developed indexing procedure and provides a list of grain orientations and positions with the corresponding assigned diffraction spots classified by $hkl$ and $\omega$. 

\section*{Analysis}

The output of the indexing process provides:

\begin{itemize}
\item The number of grains found
  \item The position of the grains found within the sample
  \item The orientation of the grains relative to each other
  \item A list of the peaks assigned to every grain
  \item The position of every peak-spot assignment
  \item The Miller indices for every peak-spot assignment
  \item The Euclidean distance for every peak-spot assignment
  \item The longest neutron wavelength of every peak-spot assignment
  \item The position and tilts of both detectors
\end{itemize}

This information can be used to evaluate the validity of the fitted grains in various ways, by computing how many peaks have been assigned to every grain or what is the mean and median assignment distance of every found grain.
Moreover, given that the peaks were cropped and stored before the indexing process, the information from the indexing can now be combined with the shape and intensity of every peak for further analysis, such as the estimation of relative volume of every grain in the sample.

In order to estimate the relative size of the grains from the diffracted intensities, we use only the diffraction data from a single family of planes. In that way, the neutron wavelengths scattered for a given Bragg angle are equivalent and therefore their intensities can be compared directly. Since the intensity of the diffraction spot is in first order proportional to the volume, we compute the average summed intensity among all spots for every different grain. The value is a relative measure of the volume of every grain, that can be used for comparison. The relative volume is calculated by

\begin{equation}
RV_i = \frac{{\cal V}_i}{\sum_{p=1}^N {\cal V}_p}
\label{eq:RelativeVolume}
\end{equation}

and

\begin{equation}
{\cal V}_i=\frac{1}{N_{i,j}}\sum_{j=1}^{N_{i,j}} I_{i,j},
\label{eq:AverageIntensity}
\end{equation}

where $RV_i$ is the relative volume of the $i^{th}$ grain, $N$ is the number of grains, ${\cal V}_i$ is the average intensity ($I_{i,j}$) of the diffraction spots ($j$) from grain $i$ restricted to a single family of planes $(hkl)$ and a narrow $\theta$ interval, $\theta \in [ \theta_1, \theta_2 ]$. 

The position and orientation of the grains (relative to each other) fitted during the indexing process can be combined with the estimated volume of the grains for the reconstruction of 3D grain maps.

\section*{Performance}

In order to test the performance of the tool under well-known conditions, synthetic (simulated) data sets with well known grain positions and orientations are used for fitting. These tests give an idea of the number of grains that can be successfully indexed by the code under realistic experimental conditions for the cubic and tetragonal symmetries. The parameters defined for the benchmarks are: 
\begin{itemize}

\item Number of grains simulated: 10
\item Detector resolution [mm]: 0.25
\item Angular range [$^{\circ}$]: 360
\item Number of angular steps (projections): 360
\item Noise introduced [\%]: 5
\item Volume of the sample [mm$^3$]: 1000
\item Sample to detector distances [mm]: 160

\end{itemize}




These parameters are set to the corresponding values unless explicitly said otherwise. We aimed to simulate that measurements often have bright spots and other artifacts, while sometimes experimental peaks are not intense enough to be segmented. For example: with a 5\% noise we remove 5\% of the simulated peaks and we add the same amount of peaks randomly placed in the synthetic data set. This is done for every grain and every $\omega$ angular step individually.

Since the position and orientation of the synthetic grains is defined \textit{a-priori}, we can evaluate the spatial and angular precision of the solver and fitting criteria chosen. Based on a detector resolution of 0.25 mm and sample-to-detector distances of 160 mm, the maximum spatial and angular deviations accepted are 0.43 mm and \ang{1} respectively.

\begin{figure}[hbt]
\centering
\includegraphics[scale=0.9]{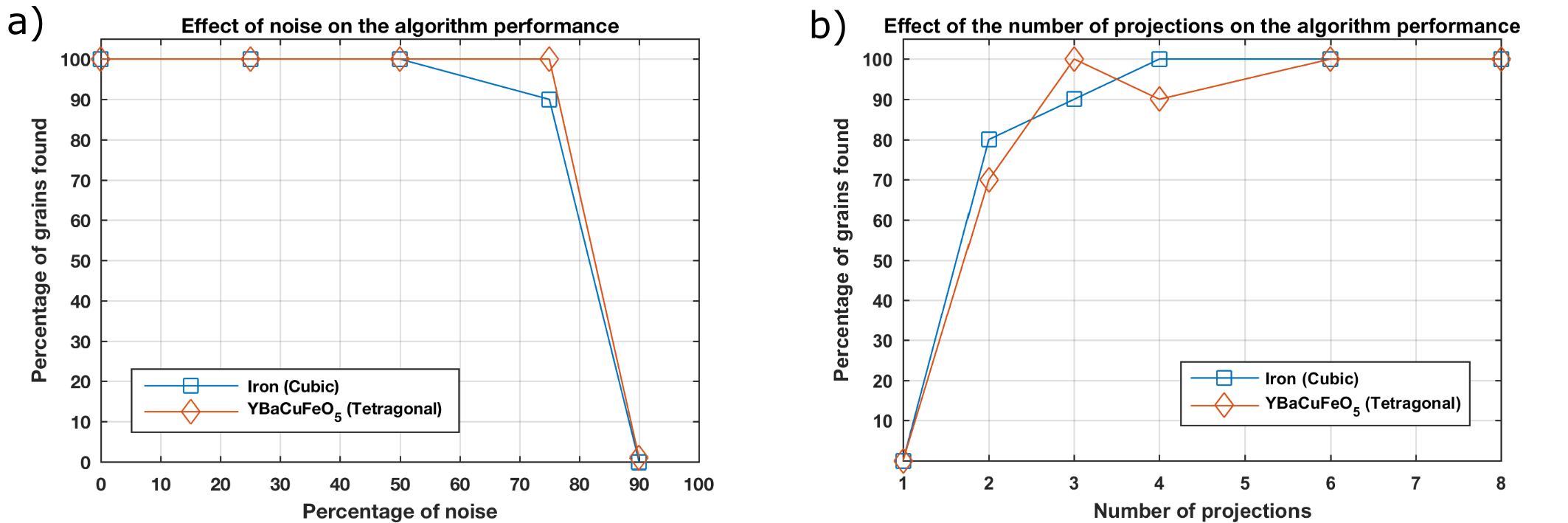}
\caption{a) Number of grains found vs percentage of spots substituted. b) Number of grains found vs number of angular steps for the measurement.}
\label{Fig:Benchmarks}
\end{figure}

As can be seen in Fig. \ref{Fig:Benchmarks}a the solver can handle up to 50\% of missing spots substituted by noise without missing any grain. Even with a rate of 75\% of peaks substituted with noise, all ten grains can be found for the tetragonal symmetry, while nine grains could be found for the cubic symmetry. When increasing the spot substitution to 90\%, no grain could be found successfully with neither the cubic nor the tetragonal symmetry. The robustness of Laue3DND with respect to noisy data sets relies on a large angular range and number of angular steps, which means better statistics. Figure \ref{Fig:Benchmarks}b shows the performance of the code with synthetic data sets using the parameters described previously, but trying to find the minimum amount of angular steps required to find 100\% of the simulated grains. For a sample with 10 grains and a data set with 360 projections around \ang{360}, only six projections are required to index the 10 grains successfully. Nevertheless, a higher number of grains in the sample would likely increase this requirement.

Finally, a test was performed following the parameters described previously, but simulating 100 grains of Fe instead of 10 within the same sample volume. Although the computation time was increased significantly up to two days, the solver was capable of successfully finding 97 grains out of 100 grains generated. Based on these results, we consider that complications arising from the indexing of a 100-grain sample might not be due to the indexing algorithm, but most likely the limitations of the watershed algorithm to deal with peak overlap.

\section*{Experimental setup}

The experiments presented in this work were performed at the E11 beam port of Helmholtz-Zentrum Berlin (HZB) with the FALCON\cite{Iles2014} Laue diffractometer installed. FALCON is composed of a back-diffraction and a forward-diffraction detector, with sizes of 400x400 mm, 4000x4000 px CCDs and a $^{6}$LiF-ZnS based scintillator with 250 $\mu$m thickness. The thickness of the scintillator screen establishes a compromise between light output and resolution \cite{Kardjilov2011}. This scintillator provides a good light output at the expense of a resolution limited to 250 $\mu$m. The thermal neutron spectrum has half of the highest intensity for the wavelengths of 0.8 \AA$ $ and 3.2 \AA. The center of the thermal neutron beam traverses the rotation axis of the sample holder, placed in between two far field detectors at 160 mm from the sample, as depicted in Fig. \ref{Fig:FALCON}a.

During the experiment, the whole sample is illuminated by the white neutron beam, diffracting simultaneously from all crystallites in the sample. Images are acquired in rotation steps of $\Delta\omega$ within the largest possible angular range (\ang{241} in our experiments) in order to provide data from as many diffraction spots (peaks) per crystallite as possible. The identification and segmentation of the individual peaks is obtained through a watershed algorithm\cite{Bala2012}. From this we obtain a list of all the peaks CMS coordinates, and a corresponding set of cropped peaks from the experimental data,  which are an ensemble of pixels showing the 2D peak profile. The list of CMS coordinates is used in the solver for the indexing procedure by comparison with the simulated spots, while the cropped peaks can be later used for analysis of the intensity.

Once the peaks CMS and profiles have been extracted, the forward model then predicts different diffraction patterns, so that the solver can find fits with the experimental data.

\section*{Experimental results}


\subsection*{Fe oligocrystal}

The first sample was a Fe oligocrystalline cylinder of 5 mm diameter and 5 mm height. The sample was measured using both forward and backward detectors with $\Delta\omega$=\ang{1} over \ang{241} and 10 seconds of exposure time per angular step.


\begin{figure}[hbt]
\centering
\includegraphics[scale=0.70]{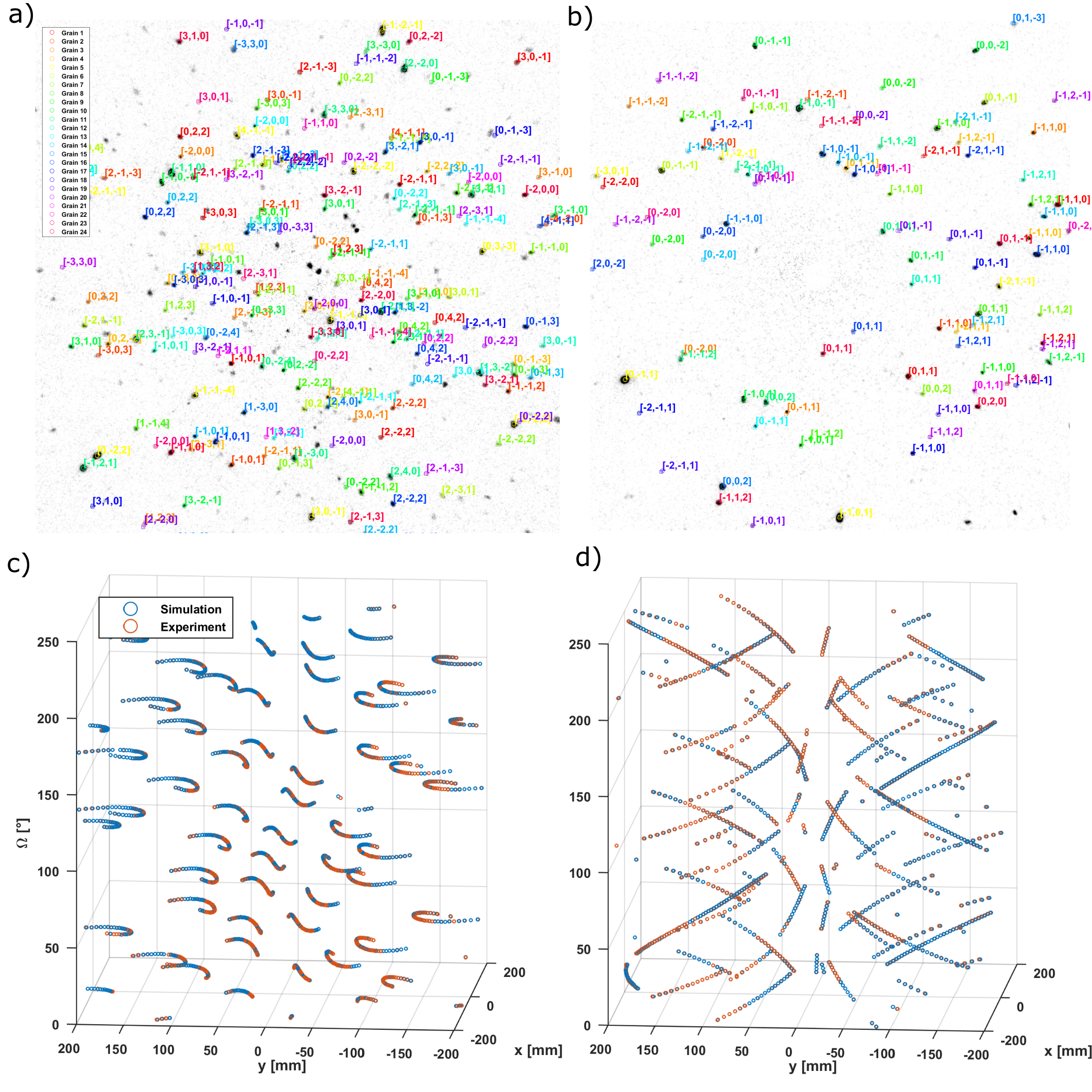}
\caption{Representation of the fitted solution against the experimental data. Top: Diffraction planes indexed in back scattering (a) and forward scattering (b) for all 24 grains. Bottom: Scatter plot with the predicted and peaks positions for every omega in back scattering (c) and forward scattering (d) for grain 12. The assignments will appear blue or orange depending on which of the two spots appears to be in front from the current viewpoint.}
\label{Fig:Fe_Spots}
\end{figure}

Figure \ref{Fig:Fe_Spots} shows two examples of backward (a) and forward (b) scattered Laue data from the oligocrystalline Fe sample. The peaks of the 24 grains found are already color coded and indexed in the image. It can be seen that some spots could not been indexed. The reason is foremost that the process for thresholding from the logarithm of the assignments will inevitably disregard some of the valid but not well fitted spots and secondly, that larger grains will generate more visible diffraction spots than smaller ones. This might only generate few visible spots from the largest structure factors for small grains, being not sufficient for the algorithm to successfully identify a grain. Hence, there might be grains in the sample yet to be found which do not pass the criteria discussed in the solver section (at least three \textit{good} assignments per angle and a median assignment distance smaller than 2 mm). Figure \ref{Fig:Fe_Spots} shows the path followed by every identified diffraction spot in back (c) and forward (d) scattering over the 241 steps of $\omega$ for a specific grain (grain 12).

\begin{table}[ht]
\centering
\begin{tabular}{l | l | l | l |l}
G  & MEA [mm] & MED [mm]  & Nspots & SOT [\%]\\
\hline

     1        &     0.61   &      0.38    &       2581            &    2.11      \\
     2         &     0.72  &       0.48   &        2504          &      2.04      \\
     3          &    0.57  &       0.38   &        2554         &       2.08      \\
     4 &             0.72  &       0.52   &        2315        &        1.89      \\
     5  &             0.60  &       0.41   &        2406       &         1.96      \\
     6   &            0.60  &       0.41   &1926              &  1.57      \\
     7   &           0.74  &       0.47   &        2236         &       1.82     \\ 
     8   &           0.55  &       0.38   &        2510         &       2.05      \\
     9   &           1.94  &       0.77   &        1313         &       1.07      \\
    10   &           0.63  &       0.44   &        2335         &       1.91      \\
    11   &          0.62  &       0.42      &      2492        &        2.00 \\
    12    &          0.73  &       0.45      &     2317         &       1.89    \\  
    13    &          0.84  &       0.51      &     2095         &       1.71      \\
    14    &          0.63  &       0.46      &     2331         &        1.9      \\
    15    &          0.58  &       0.39      &     2445         &          2.00     \\ 
    16    &          0.62  &       0.43      &     2462         &       2.01      \\
    17    &          0.95  &       0.48      &     2097         &       1.71      \\
    18    &          2.04  &       0.87      &     1121         &       0.91      \\
    19    &          2.43  &       0.97      &      887         &       0.72      \\
    20    &          0.74  &       0.42      &     2062         &       1.68      \\
    21    &          0.69  &       0.42     &      2101         &       1.71      \\
    22    &          1.25  &       0.44     &      1677         &       1.37      \\
    23    &          2.53  &       1.19     &       866         &       0.71      \\
    24    &          0.55  &       0.38     &      1974         &       1.61 \\
\hline
\bf{Total} &  & & & 40.43 \\
\end{tabular}

\caption{Results from the solver algorithm applied to the Fe oligocrystal. G: Grain number, MEA: Mean assignment distance, MED: Median assignment distance, Nspots: Number of spots assigned to peaks., SOT: Spots assigned Over Total number of peaks.}
\label{Table:1}
\end{table}

Table \ref{Table:1} provides the percentage of extracted diffraction peaks that can be assigned to the individual grains identified. The algorithm has been able to assign 40\% of the diffraction peaks (segmented with the watershed algorithm) with statistical significance. This value can be increased by lowering the criteria for a valid grain found, for instance requiring two or only one spot per scattering angle to consider a grain found, or accepting a large distance between predicted spots and experimental peaks as valid, at the cost of higher chances of making wrong assignments. Although the acceptance criteria could be less conservative, we consider that these values give enough statistical certainty to validate the indexing
method.


\begin{figure}[hbt]
\centering
\includegraphics[scale=0.8]{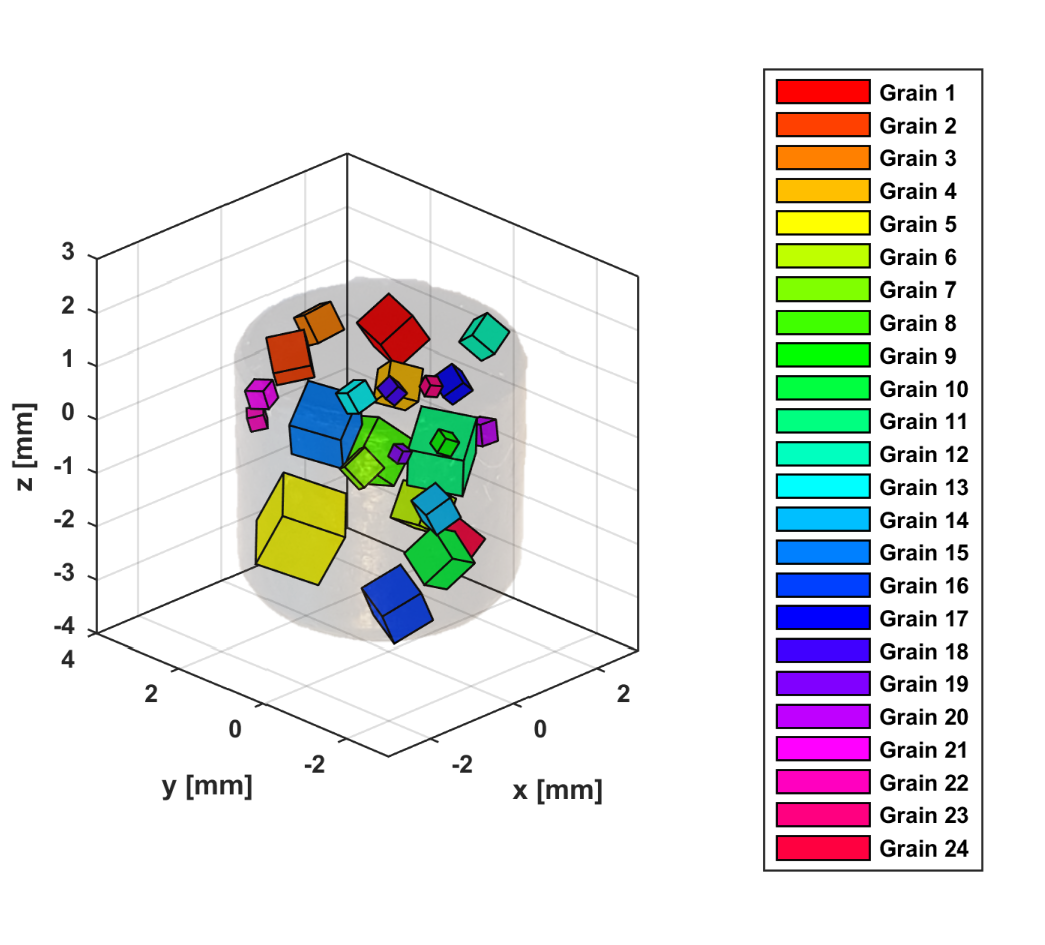}
\caption{Iron sample of 5 mm diameter and 5 mm height, with low opacity overlaid with coloured cubes representing the different grains and their and their orientations. The volume of the cubes is based on the integrated intensity of the diffraction spots of every grain and is proportional to the relative volume of the grain following equation \eqref{eq:RelativeVolume}.}
\label{Fig:Fe_Cubes}
\end{figure}

Figure \ref{Fig:Fe_Cubes} is a representation of the solution found by the solver, including the position of the CMS of every grain found in the sample and its relative size calculated from the integrated intensities of the diffraction spots using equation \eqref{eq:RelativeVolume}. Note how the positions of the grains found inside the $\alpha$-Fe sample represent a cylindrical geometry with 5 mm diameter and 5 mm height, which are the sample's dimensions. Taking into account that the boundaries for the grain CMS positions was set to 1 cm$^3$, the code was able to find all the grains inside the sample's volume without strong constraints. This underlines the accuracy of the applied method.

\subsection*{YBaCuFeO$_5$ oligocrystal}

The second sample is an oligocrystal of the high-temperature multiferroic candidate YBaCuFeO$_5$ with layered perovskite structure\cite{Morin2015,Morin2016}, in which several grains with a common $c$-axis and slightly different orientation in the $ab$ plane were formed during the process of crystal growth\cite{Morin2016a}. The main objective concerning this sample was to identify the number of grains and the respective misalignment of the $\langle$0 0 l$\rangle$ direction, as well as the relative contribution of each domain to the diffracted signal (i.e. size distribution). The sample was measured using the forward detector with $\Delta\omega$=\ang{1} over \ang{241} and 230 seconds of exposure time per angular step.

\begin{table}[ht]
\centering

\begin{tabular}{l | l | l | l |l}
G  & MEA [mm] & MED [mm]  & Nspots & SOT [\%]\\
\hline

    1 &              2.94 &        0.87 &          3200   &              7.86   \\
    2 &              4.28 &        1.04 &          1572   &              3.86   \\  
    3 &              1.77 &        0.72 &          2801   &              6.88   \\  
    4 &               1.3 &        0.69 &          3451   &              8.48   \\  
    5 &              1.45 &        0.62 &          2744   &              6.74   \\  
    6 &              1.14 &        0.55 &          1956   &              4.81   \\  
    7 &              3.11 &        0.67 &          2260   &              5.55   \\  
    8 &              0.95 &        0.67 &          4165   &             10.23   \\  
    9 &              2.46 &        0.68 &          3346   &              8.22   \\

\hline
\bf{Total} & & & & 62.63 \\
\end{tabular}

\caption{Results from the solver algorithm applied to the YBaCuFeO$_{5}$ oligocrystal. G: Grain number, MEA: Mean assignment distance, MED: Median assignment distance, Nspots: Number of spots assigned to peaks, SOT: Spots assigned Over Total number of peaks.}

\label{Table:2}
\end{table}

Table \ref{Table:2} summarizes some statistics evaluated from the fit and highlights stronger variations as compared to the previous Fe reference sample. A consideration to be taken into account in this case is the fact that the misorientation between the $\langle$0 0 l$\rangle$ directions of the different crystallites is quite small. Hence, it has to be expected that spot overlap from the reflections of this plane, more than for the others, might lead to segmentation of two actual spots as one and hence biased spot assignments.

\begin{figure}[hbt]
\centering
\includegraphics[scale=0.8]{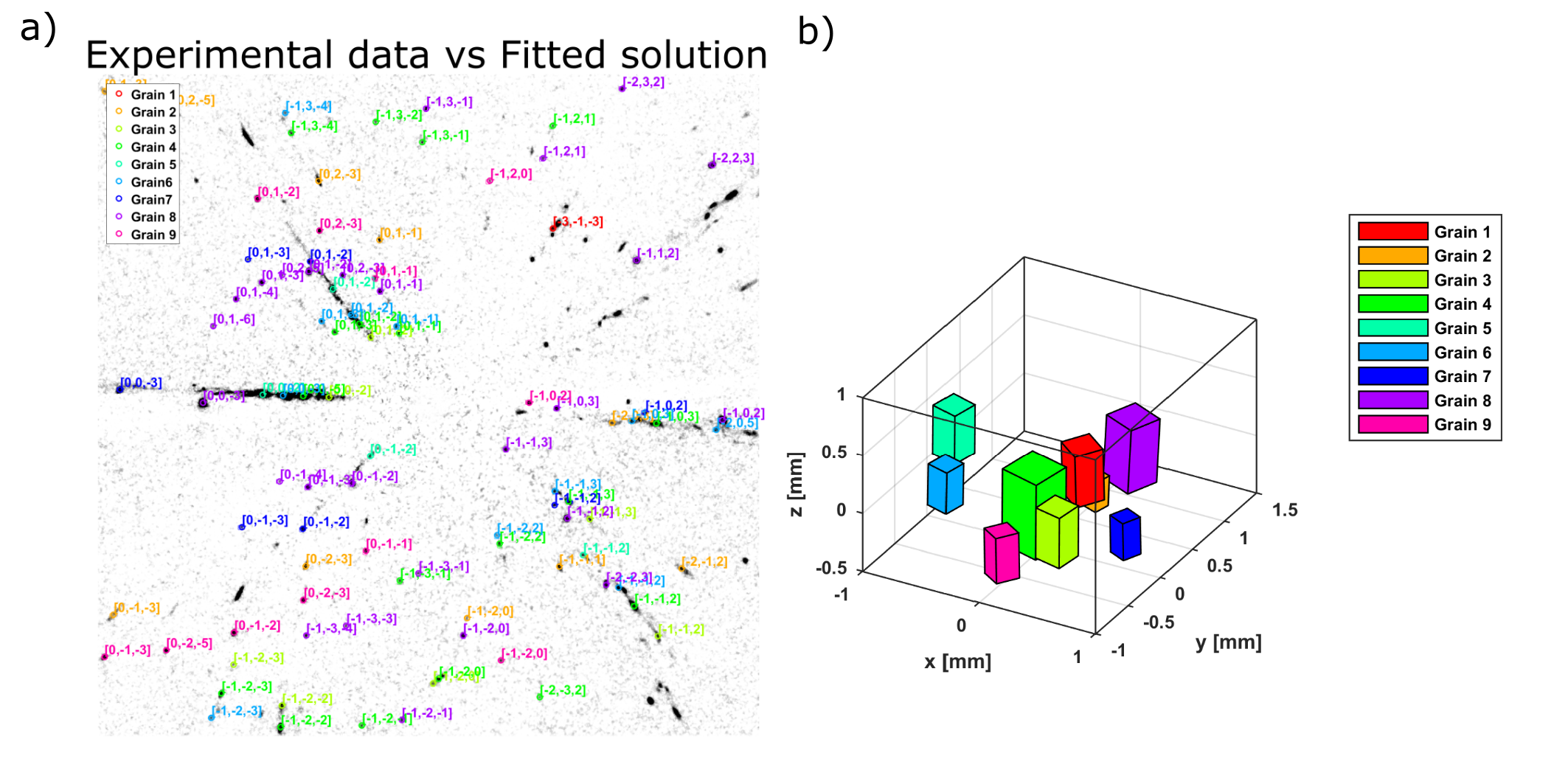}
\caption{a) Representation of the fitted solution against the experimental data for the Perovskite: Diffraction planes indexed using only the forward scattering detector. b) Plot with the 9 grains found by the algorithm inside the layered perovskite sample, represented by coloured prisms representing the different grains and their orientations. The volume of the cubes is based on the integrated intensity of the diffraction spots of every grain and is proportional to the relative volume of the grain following equation \eqref{eq:RelativeVolume}.}
\label{Fig:Perovskite_Spots_Cubes}
\end{figure}

In Fig. \ref{Fig:Perovskite_Spots_Cubes}a peaks overlap in the horizontal line around the center of the detector underline that the different grains within the sample have very similar orientation with respect to the $\langle$0 0 l$\rangle$ direction. The orientations of the two other main axis of the crystal lattice have a wider spread, generating less overlap and thus making them easier to identify and to distinguish.


Figure \ref{Fig:Perovskite_Spots_Cubes}b shows a 3D representation of the grain's positions within the sample and their relative sizes, calculated proportionally to the average integrated intensities of all diffraction spots from a single family. The box around the plotted prisms represents the search space defined as boundary conditions for the positions of the grains, which is slightly larger than the sample's size.

\section*{Discussion and Conclusions}

\textbf{Laue3DND has been presented} using experimental data for $\alpha$-Fe and YBaCuFeO$_5$ oligocrystals and corresponding synthetic data sets. This method has shown to be capable of indexing 97/100 grains on a cubic synthetic sample under realistic conditions with spatial and angular resolution of  0.43 mm and \ang{1} respectively. On experimental data sets, Laue3DND has been able to retrieve 24 grains from a $\alpha$-Fe sample and 9 grains from a YBaCuFeO$_5$. No significant differences in the code performance have been found between the cubic and tetragonal symmetries.

One of the strengths of this method is the \textbf{robustness towards noise and incomplete data sets}, as shown in Fig. \ref{Fig:Benchmarks}a, in which all the grains are found successfully even when 50\% of the peaks in the data set are substituted for random noise. A very exciting case for data sets with few projections as shown in Fig. \ref{Fig:Benchmarks}b is the possibility to generate 4D grain maps (obtaining a 3D grain map every few minutes), in order to study the evolution of the grain macrostructure of a sample under thermal or mechanical loading. 

It is important to underline the fact that this 3D grain mapping method is \textbf{several times more efficient than previous neutron methods} \cite{Peetermans2014,Cereser2017a}. Since this method does not require to select the energy of the incoming neutrons, the necessary neutron instrumentation is required and the effective neutron flux on the sample is increased between one and two orders of magnitude in comparison with other methods. In practice, this means that our new method represents a remarkable reduction in the measurements time required for neutron 3D grain mapping. The presented samples required exposure times of 10 seconds (Fe) and 215 seconds (YBaCuFeO$_5$) per angular step respectively, compared to 250 total seconds (Al\cite{Peetermans2014}) and 1 hour (Fe\cite{Cereser2017a}) for previous neutron methods. 

Given the percentage of spots which have been assigned correctly for both samples, it is expected that \textbf{some grains can still be found}. Note that the criteria used for the code to confirm the existence of a grain are based on the amount of peak-spot assignments per angle and the distance of these assignments. Since the thresholding method (Fig. \ref{Fig:Histogram}) will inevitably discard some right (and include some wrong) assignments, small grains with a low number of visible diffraction spots might not be accepted.

The current \textbf{spatial resolution} of Laue3DND (0.43 mm) is limited by the resolution of the detector itself, as well as the sample to detector distance. The experiments and simulations provided show an \textbf{orientation resolution} comparable to that of other X-ray and neutron methods (\ang{1}), arguably thanks to the far field setup, which is also a limiting factor for the spatial resolution. A first step to bring the spatial resolution closer to other neutron methods (0.1 mm) would be to set at least one of the two detectors in near field position.

The \textbf{relative volume} of the crystallites has been estimated by comparison of integrated intensities of equivalent reflections \eqref{eq:RelativeVolume}. Further quantitative analysis of the diffracted signal will be explored in future work, in order to recover the \textbf{shape of the grain boundaries}\cite{Sanchez2014,FerreiraSanchez2015} or obtain information about the wavelength distribution of the incoming beam.

The \textbf{maximum sample volume} measure with Laue3DND so far has been the $\alpha$-Fe sample with 5 mm diameter and 5 mm height. However, the FALCON beamline can accommodate samples up to 2 cm$^3$, limited by the beam diameter. Samples larger than 2 cm in one of their dimensions could be oriented vertically and translated after every rotation. Larger volumes could be analyzed at imaging beamlines using large detectors for forward scattering, but the increase in beam size will reduce the collimation of the beam and the orientation resolution
of the measurement.

The \textbf{minimum grain size} accessible with neutrons is physically limited by the coherent scattering cross section of the materials, the signal-to-noise ratio and the spatial resolution of the instrument used. We consider the practical limit for grain indexing with neutrons to be in the range of tens of micrometers for the strongest coherent nuclei, and in the order of hundreds of micrometers for others.

The \textbf{number of grains} that can potentially be indexed is harder to estimate since there are more factors in play. Peak overlap is the biggest concern when trying to index a large amount of grains, and it can be tackled by improving the angular resolution of the detection systems. One can do that by reducing the thickness of the scintillator screens at the expense of light output, in the case of scintillator based camera setups like FALCON. This should be taken into consideration for further Laue instrumentation development, since typical neutron imaging detectors already use a variety of scintillator screens suited to the resolution requirements. 



The forward model, solver and analysis tools have been so far used only to analyze neutron data, but \textbf{adapting them to X-ray Laue data} would not require fundamental changes. From the instrumentation and sample point of view, some changes are already easily implemented in the current form of the code, such as changes on the wavelength spectrum, detector geometry or sample sizes.

The study of \textbf{lower symmetry crystals} would not require fundamental changes on the Laue3DND indexing algorithms, except for monoclinic and triclinic systems. In order to index crystals from these two systems, crystal misorientation defined with quaternions would have to be coded as an alternative to the Rodrigues formulation. This is because the fundamental zone for triclinic and monoclinic symmetries corresponds to the entire Rodrigues space\cite{Randle2014}, and therefore the seeding time is infinite. Any other symmetry can be searched with the Rodrigues formulation in the present form, although the seeding time will increase with the asymmetry of the crystal.

\textbf{The code} presented in this work was written by Marc Ravent\'{o}s, S{\o}ren Schmidt and Stavros Samothrakitis for MATLAB, and can be found in the GitHub repository \textit{Laue3DND} doi: 10.5281/zenodo.1553164.


\section*{Acknowledgements}
This work was supported by the Swiss National Science Foundation, grant number 156078 and also by OP RDE, MEYS, under the project "European Spallation Source - participation of the Czech Republic - OP", Reg. No. CZ.02.1.01/0.0/0.0/16\_013/0001794. The authors want to thank Micka{\"e}l Morin for his help during the growth of the YBaCuFeO$_5$ sample. 

\section*{Author contributions}

The experiments presented in this manuscript were performed by M.R., S.Schmidt and M.T.. The YBaCuFeO$_5$ crystal was grown by M.M., T.S. and E.P.. The analysis tools were developed by S.Schmidt, M.R. and S.Samothrakitis. The manuscript was written by M.R., S.Schmidt, M.S. and C.G. All authors reviewed the manuscript.

\section*{Additional information}

\textbf{Supplementary information} accompanies this paper at: 10.5281/zenodo.1553164\\
\textbf{Competing Interests:} The authors declare no competing interests.

\end{document}